\documentstyle[12pt]{article}
\catcode`\@=11

\font\teneus=eusm10 scaled \magstep1
\font\seveneus=eusm7 scaled \magstep1
\font\fiveeus=eusm5 scaled \magstep1

\newfam\eusfam
\textfont\eusfam=\teneus  \scriptfont\eusfam=\seveneus
  \scriptscriptfont\eusfam=\fiveeus

\def\hexnumber@#1{\ifnum#1<10 \number#1\else
 \ifnum#1=10 A\else\ifnum#1=11 B\else\ifnum#1=12 C\else
 \ifnum#1=13 D\else\ifnum#1=14 E\else\ifnum#1=15 F\fi\fi\fi\fi\fi\fi\fi}

\def\Cl{\ifmmode\let\next\Cl@\else
 \def\next{\errmessage{Use \string\Cl\space only in math mode}}\fi\next}
\def\Cl@#1{{\Cl@@{#1}}}
\def\Cl@@#1{\fam\eusfam#1}

\catcode`\@=\active

\catcode`\@=11

\font\teneuf=eufm10 scaled \magstep1
\font\seveneuf=eufm7 scaled \magstep1
\font\fiveeuf=eufm5 scaled \magstep1

\newfam\euffam
\textfont\euffam=\teneuf  \scriptfont\euffam=\seveneuf
  \scriptscriptfont\euffam=\fiveeuf

\def\hexnumber@#1{\ifnum#1<10 \number#1\else
 \ifnum#1=10 A\else\ifnum#1=11 B\else\ifnum#1=12 C\else
 \ifnum#1=13 D\else\ifnum#1=14 E\else\ifnum#1=15 F\fi\fi\fi\fi\fi\fi\fi}

\def\Got{\ifmmode\let\next\Got@\else
 \def\next{\errmessage{Use \string\Got\space only in math mode}}\fi\next}
\def\Got@#1{{\Got@@{#1}}}
\def\Got@@#1{\fam\euffam#1}

\catcode`\@=\active

\catcode`\@=11

\font\tenmsx=msxm10 scaled \magstep1
\font\sevenmsx=msxm7 scaled \magstep1
\font\fivemsx=msxm5 scaled \magstep1
\font\tenmsy=msym10 scaled \magstep1
\font\sevenmsy=msym7 scaled \magstep1
\font\fivemsy=msym5 scaled \magstep1
\newfam\msxfam
\newfam\msyfam
\textfont\msxfam=\tenmsx  \scriptfont\msxfam=\sevenmsx
  \scriptscriptfont\msxfam=\fivemsx
\textfont\msyfam=\tenmsy  \scriptfont\msyfam=\sevenmsy
  \scriptscriptfont\msyfam=\fivemsy
\def\hexnumber@#1{\ifnum#1<10 \number#1\else
 \ifnum#1=10 A\else\ifnum#1=11 B\else\ifnum#1=12 C\else
 \ifnum#1=13 D\else\ifnum#1=14 E\else\ifnum#1=15 F\fi\fi\fi\fi\fi\fi\fi}

\def\Bbb{\ifmmode\let\next\Bbb@\else
 \def\next{\errmessage{Use \string\Bbb\space only in math mode}}\fi\next}
\def\Bbb@#1{{\Bbb@@{#1}}}
\def\Bbb@@#1{\fam\msyfam#1}

\catcode`\@=\active

\newcommand{\q}{q=\exp \pi i /N}
\newcommand{\ncom}{U_q(su(1,1)) }
\newcommand{\com}{U_q(su(2)) }
\newcommand{\vp}{\varepsilon}
\newcommand{\K}{{\cal K}}
\newcommand{\lp}{{\cal L}_1}
\newcommand{\lm}{{\cal L}_{-1}}
\newcommand{\hkr}{\vert h_{\mu\nu} ; kN+r \rangle}
\newcommand{\hkrp}{\vert h_{\mu\nu} ; kN+r+1 \rangle}
\newcommand{\hkrm}{\vert h_{\mu\nu} ; kN+r-1 \rangle}
\newcommand{\hk}{\vert h_{\mu\nu} ; kN \rangle}
\newcommand{\hkp}{\vert h_{\mu\nu} ; (k+1)N \rangle}
\newcommand{\hkm}{\vert h_{\mu\nu} ; (k-1)N \rangle}
\newcommand{\kr}{\vert \zeta ; k\rangle \otimes \vert j ; -j+r \rangle}

\newcommand{\V}{V_{\mu, \nu}^{irr}}
\newcommand{\zk}{\vert \zeta ; k\rangle}
\newcommand{\zkp}{\vert \zeta ; k+1\rangle}
\newcommand{\zkm}{\vert \zeta ; k-1\rangle}
\newcommand{\jr}{\vert j ; -j+r\rangle}
\newcommand{\jrp}{\vert j ; -j+r+1\rangle}
\newcommand{\jrm}{\vert j ; -j+r-1\rangle}
\newcommand{\jm}{\vert j ; m\rangle}
\newcommand{\pfun}{\Phi^\zeta}
\newcommand{\fun}{\Psi^\zeta_{j,m}(w)}

\newcommand{\po}{\phi_1^\zeta(w)}
\newcommand{\pz}{\phi_0^\zeta(w)}
\newcommand{\ph}{\phi^\zeta (w)}
\newcommand{\ps}{\psi^\zeta (w)}
\newcommand{\vc}{V_\zeta^{cl}}
\newcommand{\vq}{\mho_j}
\newcommand{\de}{\delta_\varepsilon}
\newcommand{\deo}{\delta_{\varepsilon_1}}
\newcommand{\dew}{\delta_{\varepsilon_2}}
\newcommand{\slr}{sl(2,{\Bbb R})}
\newcommand{\slc}{sl(2,{\Bbb C})}
\newcommand{\SL}{SL(2,{\Bbb R})}
\newcommand{\hxp}{\hat X_+}
\newcommand{\hxm}{\hat X_-}
\newcommand{\hxpm}{\hat X_\pm}

\begin{document}
\begin{titlepage}
\rightline{Edinburgh-/92-93/05}
\rightline{December 1993}
\vspace{1.5cm}
\begin{center}
\LARGE{More on $U_q(su(1,1))$ with $q$ a Root of Unity}  \\ [2em]
\large{Takashi Suzuki} \\[1em]
\normalsize         Department of Mathematics and Statistiics    \\
\normalsize                  University of Edinburgh            \\
\normalsize                        Edinburgh, U.K.
\end{center}
\pagestyle{plain}

\vspace{0.5cm}

\begin{abstract}
Highest weight representations of $U_q(su(1,1))$ with $q=\exp \pi i/N$
are investigated. The structures of the irreducible hieghesat weight
modules are discussed in detail.  The Clebsch-Gordan decomposition  for
the tensor product of two irreducible representations is discussed.
By using the results, a representation of $\SL\otimes\com$
is also presented  in terms of holomorphic sections which also have
$\com$ index. Furthermore we realise $Z_N$-graded supersymmetry
in terms of  the representation. An explicit realization of
$Osp(1 \vert 2)$ via the heighest weight representation of $\ncom$ with
$q^2=-1$ is  given.
\end{abstract}

\end{titlepage}

\pagestyle{plain}

\newpage

\section{Introduction}

Many works on quantum groups or $q$-deformations of universal envelopping
algebras ($q$UEA) have revealed a variety of fascinating features in both
the fields of mathematics and physics.
In particular, from the physical viewpoint, they are of great significance
with their connection to exactly solvable systems and $2D$ conformal field
theories.
The connections between $q$UEA of compact Lie algebra, $e.g.$, $\com$,
with $q$ a root of unity and rational conformal field theories
(RCFT)\cite{MoS} whose central charges are given by rational numbers have
been discussed extensively \cite{TK}-\cite{PS}.
In RCFT quantum group structures appear essentially in the monodromy
properties of conformal blocks.
The authors of Refs.\cite{GS,RRR} have shown more transparent connections
between quantum groups and RCFT by constructing good representation spaces
of $q$UEA in terms of the Coulomb gas representations of RCFT.
An important feature of the construction is that
the highest weight module of $\com$ emerges as the the family of screened
vertex operators, that is, each highest weight vector $e^j_m$ corresponds to
the screened vertex operator with $j-m$ screening operators, and the
generators of $\com$, $X_+$ and $X_-$, are represented as contour
creation and annihilation operators.
Thus, the quantum groups associated with the {\it compact} Lie algebras
when $q$ is a root of unity act as a relevant symmetry of RCFT.

In contrast, the $q$UEA, $\ncom$, of the non-compact Lie algebra, $su(1,1)$,
has not been well discussed.
Several dynamical models where $\ncom$ appears as a symmetry or a dynamical
algebra are known \cite{CEK}-\cite{KS}, and it is likely that $\ncom$ will
play a significant role in the models  with non-compact spaces.
The representation theories of $\ncom$ were given in \cite{KD}-\cite{BK}
for generic $q$, $i.e.$, $q$ not a root of unity. As in the compact case,
however, we can expect to extract new and illuminating features from $\ncom$
with $q$ a root of unity.
In Ref.\cite{MS}, Matsuzaki and the author have investigated highest weight
representations of $\ncom$ when $q$ is a root of unity and revealed a
remarkable feature: for unitary representation space,
$\ncom$ has the structure
\begin{equation}
\ncom = U(su(1,1))\otimes\com.
\end{equation}
The important point is that the non-compact nature appears through the
\lq classical' Lie algebra $su(1,1)$ and the $q$-deformed effects are only
in  $\com$.
This relation means that $\ncom$ with $q$ a root of unity is the unified
algebra of the quantum algebra $\com$ and the non-compact $su(1,1)\simeq\slr$.
The importance of this observation is recognized by noticing that the Lie
group $\SL$ has deep connection with the theory of
topological $2D$ gravity \cite{MoS}-\cite{VV}.
The spirit of this is that $\SL$ plays the role of the gauge group on the
Riemann surface $\Sigma_g$ with constant negative curvature, {\em i.e.},
with genus $g$ ($\ge2$) and the moduli space of the complex structure of
$\Sigma_g$  is identified with the moduli space of $\SL$ flat connections.
Along this line, we can expect that $\ncom$ is regarded as the symmetry of
the theory of topological gravity coupled to RCFT as a matter.
Here the $\slr$ sector of $\ncom$ gives rise to the deformation of
the complex structure of the base manifold $\Sigma_g$.
With this pespective in mind, the aim of this paper is to further investigate
$\ncom$ when $\q$ as the first step toward this goal.
We will obtain a holomorphic representation of
$\SL\otimes\com$ in terms of holomorphic vectors, which are
holomorphic sections of a line bundle over the homogeneous space $\SL/U(1)$
and that have an index with respect to $\com$.

Another novel aspect of $q$UEA at a root of unity appears in the connection
with generalized ($Z_N$-graded) supersymmetry or paragrassmann algebras
\cite{OK} as discussed in Ref.\cite{ABL}-\cite{FIK}.
Here we will present a different and concrete realization of this connection
as another remarkable consequence of the relation (1).
By using the highest weight representations, we will explicitly construct
$N$ holomorphic functions on the homogeneous space.
The $N$-th powers of the generators of $\ncom$ act on them and
give rise to infinitesimal transformations of these functions
under the transformation of the homogeneous space.
Furthermore we will see that these $N$ functions can be classified into two
sets according to their dimensions or, so called, $\slr$-spins.
One of them is the set of functions which have dimensions $\zeta$,
and the other is the set of functions with dimensions $\zeta+\frac{1}{2}$.

The organization of this paper is as follows:
In the next chapter, we briefly review the highest weight representations
of $\ncom$ in order to make this article self-contained although it has
already been done in Ref.\cite{MS}.
Chapter 3, which is the main part  of this paper, looks in detail at the
structure of $\ncom$ highest weight module when $\q$.
We also give the Clebsch-Goldan decomposition for the tensor product of two
highest weight representations.
In addition, representations of $\SL\otimes\com$ are discussed in terms of
holomorphic sections  which have  $\com$ index over the homogeneous space
$\SL/U(1)$ and discuss $Z_N$-graded supersymmetry.
In chapter 4, we explicitly derive $Osp(1\vert 2)$
via the highest weight representation of $\ncom$ with $q^2=-1$.
We conclude with some discussion.

It is convenient to summarize here the conventions and notation we will make
use in this paper; ${\Bbb Z}_+$ stands for the non-negative integers, $i.e.$,
${\Bbb Z}_+ = \{0, 1, 2, \cdots\}$.
$[n]$ for $n\in {\Bbb Z}$ describes a $q$-integer defined by
$[n]=(q^n-q^{-n})/(q-q^{-1})$.
This convention is useful in our later discussion because, for
${}^\forall n \in {\Bbb Z}$,  $[n] \in {\Bbb R}$
if  $q \in {\Bbb R}$ or $\vert q \vert =1$.
Finally $\left[ \begin{array}{c} n \\ r \end{array} \right]_q = [n]!/
[n-r]![r]!$
is a $q$-binomial coefficient.


\section{Highest Weight Representations of $U_q(su(1,1))$}

To begin with, it is helpful for our discussions to give a brief
review of unitary representations of the classical
Lie algebra $su(1,1)$. This appears as a non-compact real form of the
Lie algebra $\slc$ generated by $E_+, E_-$ and $H$.
The relations among them are
\begin{equation}
  [E_+, E_-] = 2H, \quad\quad [H, E_\pm] = \pm E_\pm
\end{equation}
The difference between the compact real form $su(2)$ and the non-compact one
$su(1,1)$ will appear only through the definitions of Hermitian conjugations.
They are  defined by,
\begin{equation}
\begin{array}{lll}
E_\pm^\dagger =  E_\mp,  &\quad\quad H^\dagger = H,
                             &\quad\quad\quad{\rm for}\quad su(2), \\
E_\pm^\dagger = - E_\mp, &\quad\quad H^\dagger = H,
                             &\quad\quad\quad{\rm for}\quad su(1,1).
\end{array}
\end{equation}
Via the substitution  $E_\pm\rightarrow \mp G_\mp, \, E_0\rightarrow G_0$
we can get another formulation of $su(1,1)\simeq \slr$.
Now  the relations and Hermitian conjugations are,
\begin{equation}
  \begin{array}{ll}
       [G_n, G_m]=(n-m)G_{n+m},   &  n, m=0, \pm 1,   \\
   \rule{0mm}{.7cm}
       \quad G_\pm^\dagger = G_\mp, \quad  G_0^\dagger =G_0,
            & {\rm for} \quad su(1,1),  \\
  \rule{0mm}{.5cm}
       \quad G_\pm^\dagger = -G_\mp, \quad G_0^\dagger =G_0,
            & {\rm for} \quad su(2)
   \end{array}  \label{eq:another}
\end{equation}

\noindent
We will use the latter formulation in chapter 3.

Representations are classified by means of eigenvalues of Cartan operator
$H$ and the second Casimir operator.
It is well known that there are four classes of unitary irreducible
representations: (a) Identity $I$; The trivial representation of the form
$I= \{ \vert 0 \rangle \}$.
(b) Discrete series ${\cal D}_n^+$ ; ${\cal D}_n^+ = \{\vert k+ \phi \rangle
\;\vert\; k = n, n+1, \cdots \}$ with $n \in {\Bbb Z}_+$, that is,
each representation ${\cal D}_n^+$
is bounded below such that $E_- \vert n+ \phi \rangle = 0$.
The state $\vert n + \phi \rangle$ is called a wighest weight state.
${\cal D}_n^+$ is refered to as the highest weight representation and
we will concentrate only on this type.
(c)  Discrete series ${\cal D}_n^-$ ; ${\cal D}_n^- = \{\vert k-\phi \rangle
\; \vert \; k = -n, -n-1, \cdots\}$ for  $n \in {\Bbb Z}_+$, that is,
each representation ${\cal D}_n^-$
has the upper bound state such that $E_+ \vert -n- \phi \rangle = 0$.
This type of representation is called the lowest weight representaion.
(d) Continuous series ${\cal B}$ : Representations of the form
${\cal B} = \{ \vert k + \phi \rangle \; \vert \; k \in {\Bbb R} \}$.
In the cases (b)$\sim$(d),  $\phi$ takes its value in $0<\phi\le 1$ and
it cannot be further determined.
In particular, representations ${\cal D}_n^\pm$ are
not really discrete in this sense.
Only by a consideration of the Lie group action of $SU(1,1)$ on the highest
or lowest weight representations, does the discreteness arise.
Then $\phi$ is determined to be $1/2$, or 1.
It will turn out that the highest weight represaentation of $\ncom$ with
$q$ a root of unity is actually a discrete series without any other
considerations. Let us proceed to the quantum cases.


\subsection{The case when $q$ is not a root of unity}

We briefly summarize the highest weight representation of the non-compact real
form of $U_q(\slc)$ when $q$ is not a root of unity in order to make
the differences from the case $q$ when is a root of unity clear.
Generators and relations of $U_q(\slc)$ are as follows.

\vspace{.3cm}

\noindent
{\bf Definition 2.1}$\quad$  $U_q(\slc)$ is generated by $X_+,\,
X_-,\, K$ with relations among them given by
\begin{equation}
[X_+, X_-] = \frac{K^2-K^{-2}}{q-q^{-1}}, \quad   KX_\pm = q^{\pm1} X_\pm K.
\label{eq:relation}
\end{equation}

\vspace{1.0cm}
\noindent
As in the classical case, the Hermitian conjugation rule determines whether
its real form is  compact or non-compact.
Conjugations consistent with the relations
exist if and only if $q$ is real or $\vert q\vert =1$.
The non-compact real form  $U_q(su(1,1))$  can be obtained by defining
the conjugation as follows:
\begin{equation}
\begin{array}{clll}
{}& K^\dagger = K, &\quad X_\pm^\dagger = -X_\mp, &\quad \quad
   {\rm when} \; q\in {\Bbb R} \\
{}& K^\dagger=K^{-1}, &\quad X_\pm^\dagger = -X_\mp, &\quad \quad
   {\rm when} \; \vert q\vert =1
\end{array}
\label{eq:qhermite}
\end{equation}
If we take the conjugation of $X_\pm$ to be $X_\pm^\dagger = X_\mp$ instead of
the above, we obtain compact real form, $U_q(su(2))$.
A Hopf algebraic structure results upon defining a coproduct $\Delta$,
a counit $\epsilon$ and an antipode $\gamma$ by
$\Delta(K) = K \otimes K, \; \Delta(X_\pm) = X_\pm \otimes K^{-1} + K \otimes
X_\pm$, $\epsilon(K) = 1, \; \epsilon(X_\pm) = 0$,
$\gamma(K) = K^{-1}, \; \gamma(X_\pm) = -q^{\mp1} X_\pm$.

One can represent $U_q(su(1,1))$
by constructing a highest weight module as in the classical case.
Each module is characterized by a positive parameter $h$
which is the highest weight.  The highest weight $h$ is specified
by the second order Casimir operator. The highest weight module over a
highest weight vector $\vert h; 0 \rangle$ is given by

\begin{equation}
V_h=\{\vert  h; r\rangle \; \vert\;
\vert  h; r\rangle := \frac{(X_+)^r}{[\,r\,]!} \vert  h; 0\rangle, \;
 r\in {\Bbb Z}_+\}
\end{equation}
where the highest weight vector is characterized by

\begin{equation}
X_-\vert  h; 0\rangle =0,
          \quad K\vert  h; 0\rangle = q^h \vert  h; 0\rangle.
\end{equation}
Using the definition of weight vectors $\vert  h; r\rangle$ and the relations
(\ref{eq:relation}), the action of $U_q(su(1,1))$ on the highest weight
module $V_h$ is given as follows,
\begin{eqnarray}
X_+\vert  h; r\rangle &=& [r+1] \vert h; r+1 \rangle, \label{eq:xpaction} \\
X_-\vert h; r\rangle &=&  -[2h +r -1] \vert h; r-1 \rangle,
                                           \label{eq:xmaction}\\
K \vert h; r\rangle &=& q^{h+r} \vert h; r\rangle. \label{eq:kaction}
\end{eqnarray}
With the Hermitian conjugation (\ref{eq:qhermite}),
the norm of the state $\vert h; r \rangle$ is given as the inner product
$\parallel \vert h;r \rangle \parallel^2 :=\langle h;r\vert h;r\rangle$
and is
\begin{equation}
\parallel \vert h;r \rangle \parallel^2 =  \left[
       \begin{array}{c} 2h+r-1 \\ r \end{array}
         \right]_q,  \label{eq:norm}
\end{equation}
where we have normaized the norm of the highest weight vector as
$\parallel \vert h; 0  \rangle \parallel^2 =1.$
Since the $q$-integer $[2h+r-1]$ in the r.h.s. of eq.(\ref{eq:xmaction})
is always non-zero for $r\ge 1$, we see that no other submodules appear,
that is,  the highest weight module $V_h$ is irreducible.
Furthermore, if $q \in {\Bbb R}$, we can also see that all weight vectors have
positive definite norm, $i.e.$, $V_h$ is unitary.
Notice that every highest weight module $V_h$ is  isomorphic to
the highest weight module $V_h^{cl}$ of the classical $su(1,1)$.


\subsection{The case when $q$ is a root of unity}

In the following we set the deformation parameter $q$ to be
$q=\exp \pi i /N$. Then $q^N=-1$ and $[N]=0$.
In this case the situation is drastically different from both the classical
case and the case with generic $q$  and some difficulties appear.

The first problem is that the norm of the $N$-th state diverges due to the
$[N]$ in the denominator of eq.(\ref{eq:norm}).
The same problem occues in the $U_q(su(2))$ case as well.
Fortunately in that case the problem could be resolved utilizing to the fact
that highest  and  lowest weight states exist for each module.
Namely, we can always choose a value of  the highest weight
such that the module does not have the state whose norm diverges.
This is the origin of the finiteness of the number of the highest weight
states in the unitary representation of $U_q(su(2))$.
However in the non-compact case, we cannot remedy this problem in such a
manner because the unitary representation is always of infinite dimension.
Instead, in order for the $N$-th state to have finite norm, we have to impose
the condition that there exists an integer $\mu$ satisfying
\begin{equation}
   [ 2h + \mu - 1 ] = 0.
\end{equation}
This factor appearing in the numerator cancells the factor $[N]$ at the
$N$-th state. Of course, it is neccessary that $\mu\le N$.
Thus we are led to the following proposition:

\vspace{.3cm}
\noindent
{\bf Proposition 2.2}$\quad$
A highest weight module
of $U_q(su(1,1))$ with $q=\exp \pi i / N$ is well-defined if
and only if the highest weight is labelled by two integers $\mu$ and $\nu$ as

\begin{equation}
  h_{\mu\nu} =\frac{1}{2} (N\nu -\mu +1), \quad\quad \mu= 1, 2, \cdots N,
  \quad \nu \in {\Bbb N}.  \label{eq:hmn}
\end{equation}

\vspace{.5cm}
\noindent
The restriction $\nu\in {\Bbb N}$ (equivalently $h_{\mu\nu}>0$) follows
because we are considering the highest weight representations.
The important point coming from the proposition is that the highest
weight representations of $U_q(su(1,1))$ with $\q$ are actually a
discrete series with the highest weights takeing values in
$\{\frac{1}{2}, 1, \frac{3}{2}, \cdots\}$.
Unlike the classical case, no consideration about a group action is
needed to show this discreteness.
As we will see in section 3.2, these values of $h_{\mu\nu}$ are compatible
with representations of $\SL\otimes\com$.

In the construction of a highest weight module
we come upon further difficulties.
First, noticing (\ref{eq:xpaction}),(\ref{eq:xmaction}),
the generators $X_\pm$ are nilpotent on the module
due to  $[N]=0$ and $ [ 2h_{\mu\nu} + \mu - 1 ] = 0$, {\em i.e.},

\begin{equation}
(X_\pm)^N \vert \psi\rangle =0,
\end{equation}
where $\vert\psi\rangle$ is an arbitrary state in the highest weight module
on $\vert h_{\mu\nu};0\rangle$. Therefore one cannot move from a state to
another state by acting $X_+$ or $X_-$ successively.
However $(X_\pm)^N / [N]!$ is well-defined on the module,
and one can reach every stste with the set $\{X_\pm, (X_\pm)^N/[N]!\}$.
Secondly, the operator  $K$ is not sufficient to specify
the  weight of a state due to the relation $K^{2N}=1$.
Indeed, two states $\vert h; r \rangle$ and $\vert h; r+2N \rangle$ have
the same eigenvalue $q^{h_{\mu\nu}+r}$ with respect to the operator $K$.
This fact means that we need other operator in addition to $K$ to specify
weight completely.
These problems have already appeared in the compact case and Lusztig resolved
them by adding generators and redefining $U_q(\slc)$ \cite{Lu}.
His method is applicable to our non-compact case as well, and so
we add new generators

\begin{equation}
L_{1}:= -\frac{(-X_-)^N}{[N]!}, \quad
L_{-1}:= \frac{(X_+)^N}{[N]!}, \quad
      L_0:=\frac{1}{2}
             \left[ \begin{array}{c}  2H+N-1 \\
                                         N         \end{array} \right]_q,
\label{eq:ldef}
\end{equation}
where $K=q^H$.
In order to obtain non-compact representations,
Hermitian conjugations are  given by the second line of (\ref{eq:qhermite})
for the operators $X_\pm, K$ and, therefore, those for the new operators are
\begin{equation}
L_{\pm1}^\dagger = -L_{\mp1}, \quad L_0^\dagger = L_0.
\end{equation}

Now we are ready to investigate the highest weight representations of
$U_q(su(1,1))$ with $q$ a root  of unity.
Let us construct the highest weight module,  $V_{\mu,\nu}$,
on the highest weight vector $\vert h_{\mu\nu}; 0 \rangle$,

\begin{equation}
   V_{\mu,\nu}= \{ \vert h_{\mu\nu}; r\rangle \>\vert \> r\in {\bf Z}_+\}.
\end{equation}
The highest weight vector is characterized by
\begin{eqnarray}
      X_-\vert h_{\mu\nu}; 0 \rangle &=&
              L_{-1} \vert h_{\mu\nu}; 0 \rangle = 0,  \\
      K \vert h_{\mu\nu}; 0 \rangle
              &=& q^{h_{\mu\nu}} \vert h_{\mu\nu}; 0 \rangle, \quad
         L_0 \vert h_{\mu\nu}; 0 \rangle = \ell\, \vert h_{\mu\nu}; 0 \rangle,
\end{eqnarray}
where $\ell := \frac{1}{2} \left[ \begin{array}{c}
                                   2h_{\mu\nu} +N -1 \\
                                                N
                      \end{array} \right]_q $.
Although this $q$-binomial coefficient includes a factor $\frac{0}{0}$,
we can estimate it by using  $[kN] / [N] = (-)^{k-1} k$ and obtain
$\ell = (-)^{N\nu+\mu}\frac{1}{2}\nu$.
In the discussions of the highest weight module, it is convenient to consider
two cases  (I) $1 \le \mu \le N-1$ and (II) $\mu = N$ separately,
since zero-norm states appear only in the case (I).

\vspace{0.3cm}
\noindent
(I) $1\le \mu\le N-1$: $\quad$
The drastic difference from both the classical and generic cases is in the
fact that the  highest weight module is not of itself irreducible owing to the
state $\vert h_{\mu\nu}; \mu \rangle$.
By the definition of the parameter $\mu$, this state has zero norm.
The appearence of the zero-norm state is indispensable for obtaining
a well defined representation.
The state is not only a zero-norm state  but also a highest weight state
(we will call such a state a null state) due to the relations,

\begin{equation}
   X_- \vert h_{\mu\nu}; \mu \rangle = 0, \quad\quad
   L_{-1} \vert h_{\mu\nu}; \mu \rangle = 0.
\end{equation}
The first equation comes from the definition of $\mu$,
and the second equation follows from the fact that there is not the
corresponding state in $V_{\mu,\nu}$ because $\mu \le N-1$.
The weight of $\vert h_{\mu\nu}; \mu \rangle$  is
$h_{\mu\nu}+\mu = h_{-\mu\nu}$ and therefore the state
$\vert h_{\mu\nu}; \mu \rangle$  can be regarded as the highest weight
state $\vert h_{-\mu\nu}; 0 \rangle$.
Thus the original highest weight module $V_{\mu,\nu}$
has the submodule $V_{-\mu,\nu}$ on the null state
$\vert h_{\mu\nu}; \mu \rangle$ and, therefore,
the module $V_{\mu,\nu}$ is not irreducible.
Further, we can easily show that  $V_{-\mu,\nu}$ has again the submodule
$V_{\mu,\nu+2}$ which has the submodule $V_{-\mu,\nu+2}$ and so on.
Finally we obtain the following embedding of the submodules in the original
highest weight module $V_{\mu,\nu}$;
\begin{equation}
V_{\mu,\nu}\rightarrow V_{-\mu,\nu}\rightarrow V_{\mu,\nu+2}\rightarrow
V_{-\mu,\nu+2}\rightarrow \cdots \rightarrow V_{\mu,\nu+2k}\rightarrow
V_{-\mu,\nu+2k} \rightarrow\cdots
\end{equation}
An irreducible highest weight module $V_{\mu,\nu}^{irr}$ on the highest
weight state $\vert h_{\mu\nu}; 0 \rangle$ is constructed by subtracting
all the submodules, and finally we obtain
\begin{equation}
V_{\mu,\nu}^{irr} = \sum_{k \in {\bf Z}_+} V_{\mu,\nu}^{(k)},
\label{eq:block}
\end{equation}
where $V_{\mu,\nu}^{(k)} := V_{\mu,\nu+2k} - V_{-\mu,\nu+2k}$.
Notice that all the zero-norm states that lie on the
levels from $(kN+\mu)$ to $((k+1)N-1)$ disappear by the subtraction.
The remarkable point is that,  the irreducible highest weight module
has {\it block structure}, that is to say,
$V_{\mu,\nu}^{irr}$ consists of infinite series of blocks
$V_{\mu,\nu}^{(k)}, \;k=0, 1, \cdots$, and each block has finite number of
states, $\vert h_{\mu\nu}; kN+r \rangle, \; r=0, 1, \cdots, \mu-1$.
The operators $X_\pm$ move states in each block according to
(\ref{eq:xpaction}), (\ref{eq:xmaction}) together with the conditions
$X_-\vert h_{\mu\nu}; kN \rangle=X_+\vert h_{\mu\nu}; kN+\mu-1 \rangle=0$.
On the other hand, the operators $L_{\pm1}$ map the state
$\vert h_{\mu\nu}; kN+r \rangle$ to $ \vert h_{\mu\nu}; (k\mp1)N+r \rangle. $

\vspace{0.3cm}
\noindent
(II) $\mu =N$: $\quad$
In this case, no null state  appears in $V_{N,\nu}$.
The module is, therefore, irreducible of itself.
The only difference from the classical and generic cases is that
the actions of $X_+$ and $X_-$, respectively, on the $(kN-1)$-th
state and the $kN$-th state vanish, {\em i.e.},
$X_+\vert h_{N,\nu};kN-1 \rangle=0$ due to $[N]=0$
and $X_-\vert h_{N,\nu};kN \rangle=0$ due to $[2h_{N,\nu}+N-1]=0$.
Instead, the state $\vert h_{N,\nu};kN \rangle$ can be generated by the
operation of $L_{-1}$ on the state $\vert h_{N,\nu};(k-1)N \rangle$.
Thus the irreducible highest weight module also has the {\it block structure}
as in the case (I),

\begin{equation}
 V_{N,\nu}^{irr} = \sum_{k \in {\bf Z}_+} V_{N,\nu}^{(k)},
\label{eq:blockb}
\end{equation}
where $V_{N,\nu}^{(k)}$ is the $k$-th block which consists of
$\vert h_{N\nu}; kN+r\rangle, \; r=0, 1, \cdots, N-1.$

Now we have obtained irreducible highest weight modules of $\ncom$ when $\q$.
The crucial point is that unlike the classical or generic cases,
every irreducible module has the block structure (\ref{eq:block}) or
(\ref{eq:blockb}) which can be written as $V_k \otimes V_r$, where $V_k$
is the space which consists of infinite number of blocks
and $V_r$ stands for a block having
finite number of states $\vert h_{\mu\nu} : kN+r\rangle, \; r=0 \sim \mu-1$.
It will turn out that the block structure is the very origin of novel
features of $\ncom$ at a root of unity.
Furthermore we should notice that the irreducible module $\V$ is not
neccessarily unitary, because $[ x ]$ for a positive integer $x$
is not always positive.

Finally we present character formula of the representation.
\begin{eqnarray}
\chi_{\mu\nu}(x) &:=&  {\rm Tr}_{\V} x^H = \sum_{k=0}^\infty
\left( \frac{x^{h_{\mu\,\nu+2k}}}{1-x} -
\frac{x^{h_{-\mu\,\nu+2k}}}{1-x}\right)  \nonumber \\
&=&\frac{x^{h_{\mu\,\nu}}}{1-x} \frac{1 - x^\mu}{1 - x^N}.
\label{eq:character}
\end{eqnarray}
This holds for $1\le\mu\le N$.
In the $\mu=N$ case, the character $\chi_{N\nu}(x)$ is the same as that
of classical highest weight representation of $su(1,1)$.


\section{Structure of $U_q(su(1,1))$ with $\q$}

The aim of this chapter is to elaborate on the irreducible highest
weight module $V_{\mu,\nu}^{irr}, \; \mu=1, 2, \cdots N$ and its attendant
the block structure.
It is convenient to write the level of a state in $V_{\mu,\nu}^{irr}$
by $kN+r$ with $r=0, 1, \cdots, \mu-1$ rather than $r = 0, 1, \cdots$,
and we always adopt this notation hereafter.


\subsection{Main Theorem}

In this section we prove the following theorem which states the structure of
the irreducible highest weight module of $\ncom$.
%
%

\vspace{.7cm}

\noindent
{\bf Theorem 3.1}$\quad$
When $\q$, the irreducible highest weight $U_q(su(1,1))$-module is isomorphic
to a tensor product of two spaces as  follows;
\begin{equation}
V_{\mu,\nu}^{irr}  \simeq V_\zeta^{cl} \otimes \mho_j,
\end{equation}
where $\zeta=\frac{1}{2}\nu$ and  $j=\frac{1}{2}(\mu-1)$.
These two spaces $V_\zeta^{cl}$ and $\mho_j$
are understood as follows;
\begin{tabbing}
(1) if $\V$ \= is unitary (see proposition 3.2 below), then \\
            \> $V^{cl}_\zeta$ :
              unitary irreducible infinite dimensional $su(1,1)$-module  \\
            \>$\mho_j$ :
              unitary irreducible finite dimensional $U_q(su(2))$-module.
\end{tabbing}

\vspace{0.1cm}
\noindent
\begin{tabbing}
(2) \= if $\V$ \=is not unitary, there are three cases as follows;  \\
    \> (2-1) $\nu\,\in 2{\Bbb N}+1$ and $\nu N+\mu\,\in 2{\Bbb N}+1$\\
    \>             \> $V^{cl}_\zeta$ :
        non-unitary irreducible infinite dimensional $su(2)$-module  \\
    \>             \> $\mho_j$ :
             unitary irreducible $U_q(su(2))$-module.  \\
    \> (2-2) $\nu\,\in 2{\Bbb N} $ and $\nu N+\mu\,\in 2{\Bbb N} $  \\
    \>             \> $V^{cl}_\zeta$ :
          unitary irreducible infinite dimensional $su(1,1)$-module  \\
    \>             \> $\mho_j$ :
       non-unitary irreducible  finite dimensional $U_q(su(1,1))$-module. \\
    \> (2-3) $\nu\,\in 2{\Bbb N} $ and $\nu N+\mu\,\in 2{\Bbb N}+1$  \\
    \>             \> $V^{cl}_\zeta$ :
          non-unitary irreducible infinite dimensional $su(2)$-module  \\
    \>             \> $\mho_j$ :
        non-unitary irreducible  finite dimensional $U_q(su(1,1))$-module.
\end{tabbing}

\vspace{0.5cm}
\noindent
To prove this  theorem we first find an isomorphism $\rho$ between
$\V$ and $V_k\otimes V_r$, and then  discuss what $V_k$ and $V_r$ are.
(For the time being, we will denote the two spaces
$V^{cl}_\zeta$ and $\mho_j$ in the theorem respectively
by $V_k$ and $V_r$ in accordance with the previous chapter.)
We begin with the observation that the operator
$(X_\pm)^{kN+r}/[kN+r]!$ can be rewritten in terms of $X_\pm$ and $L_{\pm1}$.
By the straightforward calculation, one finds

\begin{equation}
\frac{X_\pm^{kN+r}}{[kN+r]!} = (-)^{\frac{1}{2} k(k-1)N+kr}
\frac{X_\pm^r}{[r]!}\frac{(L_{\mp1})^k}{k!}.
\end{equation}
Furthermore one should notice that the genearators $L_n, \;n=0, \pm1$
(anti-)commute with the generators $X_\pm, K$ on the module
$V_{\mu, \nu}^{irr}$.
That is, for $ \vert \psi \rangle \in V_{\mu,\nu}^{irr}$,
\begin{equation}[X_\pm, L_n]\vert \psi \rangle =0,
\quad K^{-1} L_n K\vert \psi \rangle =-L_n \vert \psi \rangle,
\quad\quad n=0, \pm1,
\end{equation}
These equations mean that up to a sign we can reach the $(kN+r)$-th state by
letting  $L_{-1}^k/ k!$  and $X_+^r/[r]!$ operate separately
on the highest weight state.
These facts indicate  that there exits a map $\rho\; ; \;
\V \rightarrow V_k \otimes V_r$ and the map $\rho$ induces another map
$\hat \rho\; : \; U_q(su(1,1)) \rightarrow U_k \otimes U_r$
where $U_r$ and $U_k$ are the universal enveloping algebras generated
by $X_\pm, K$ and  $L_n, \; n=0, \pm1$, respectively.
In order to define the maps $\rho$ and $\hat\rho$,
we observe the actions of $X_{\pm}, K$ and $L_n$ on $V_{\mu, \nu}^{irr}$.
The actions of $U_r$ are generated by
\begin{eqnarray}
X_+ \hkr &=& (-)^k [r+1] \hkrp, \label{eq:qxpaction}\\
X_- \hkr &=& -(-)^{\nu+1}(-)^k [\mu - r] \hkrm, \label{eq:qxmaction}\\
K \hkr &=& (-)^{\frac{\nu}{2} +k} q^{-j+r} \hkr, \label{eq:qkaction}
\end{eqnarray}
and as far as the operators $L_n$ are concerned, it is sufficient to consider
the actions on $\hk$. They are as follows;

\begin{eqnarray}
L_1 \hk &=& (-)^{kN} (k+1) \hkp, \label{eq:lpaction}\\
L_{-1} \hk &=& -(-)^{\nu N + \mu}(-)^{kN} (\nu+k-1) \hkm, \label{eq:lmaction}\\
L_0 \hk &=& (-)^{\nu N+\mu} \left( \frac{1}{2} \nu +k \right) \hk.
                  \label{eq:lzaction}
\end{eqnarray}
The map $\rho$ can be defined as follows,

\begin{equation}
\rho (\hkr) = (-)^{\frac{1}{2}k(k-1)N+kr}\kr, \label{eq:defiso}
\end{equation}

\noindent
where $\zeta=\frac{1}{2}\nu$ and $j=\frac{1}{2}(\mu-1)$.
This is an isomorphism between $\V$ and $V_k \otimes V_r$.
{}From (\ref{eq:qxpaction})-(\ref{eq:lzaction}) together with (\ref{eq:defiso})
we obtain the map
$\hat\rho$: $\ncom\rightarrow U_k\otimes U_r$, such that
$ \rho ({\cal O} \vert\psi\rangle)
= \hat\rho({\cal O})\rho(\vert\psi\rangle)$ for $\vert\psi \rangle
\in\V, \; {\cal O}\in \{X_\pm, K, L_{0, \pm1}\}$.
Explicitly we find
\begin{eqnarray}
&& \begin{array}{l} \hat\rho(X_+) = {\bf 1}  \otimes (-)^{\nu+1} J_+, \\
\quad\quad\quad\quad\quad \mbox{where}\quad
             J_+ \jr = (-)^{\nu+1} [r+1]\jrp, \end{array} \label{eq:defxp} \\
&&  \begin{array}{l} \hat\rho(X_-) = {\bf 1} \otimes (-J_-), \\
\quad\quad\quad\quad\quad
        \mbox{where}\quad J_-\jr = (-)^{\nu+1}[2j-r+1]\jrm, \end{array}
                                                           \label{eq:defxm}\\
&& \begin{array}{l} \hat\rho(K) = (-)^{L_0}{\bf 1} \otimes \K, \\
\quad\quad\quad\quad\quad \mbox{where}\quad \K \jr = q^{-j+r}\jr \end{array}
                                                         \label{eq:defk}\\
&& \begin{array}{l} \hat\rho(L_1) = (-G_1)
                                   \otimes {\bf 1}, \\
\quad\quad\quad\quad\quad  \mbox{where}\quad
G_1\zk = (-)^{\nu N+\mu} (2\zeta+k-1)\zkm, \end{array}   \label{eq:deflp}\\
&& \begin{array}{l} \hat\rho(L_{-1}) = (-)^{\nu N + \mu} G_{-1}
                                  \otimes {\bf 1}, \\
\quad\quad\quad\quad\quad  \mbox{where}\quad
            G_{-1} \zk = (-)^{\nu N + \mu}(k+1) \zkp, \end{array}
                                            \label{eq:deflm}\\
&& \begin{array}{l} \hat\rho(L_0) = (-)^{\nu N + \mu}G_0 \otimes {\bf 1}, \\
\quad\quad\quad\quad\quad  \mbox{where}\quad G_0\zk= (\zeta +k)\zk,
                        \end{array} \label{eq:deflz}
\end{eqnarray}
Here we have included some sign factors $(-)^{\nu+1}, \; (-)^{\nu N+\mu}$
for later convenience.
The next step to complete the proof of the theorem is to examine in detail the
modules $V_k$ and $V_r$ which are spanned by $\zk$ and $\jr$, respectively.
We calculate the commutation relations among $J_+, J_-, \K$
and among $G_1, G_{-1}, G_0$.
As for $V_k$ and $U_k$, one finds from the equations
(\ref{eq:deflp})-(\ref{eq:deflz}) that,
\begin{equation}
  [G_n, G_m] = (n-m) G_{n+m}, \quad\quad\quad n,m=0, \pm1.
\end{equation}
They are just the relations in (\ref{eq:another}) of the classical $\slc$
and we set the Hermitian conjugations as
\begin{equation}
G_{\pm1}^\dagger= (-)^{\nu N + \mu} G_{\mp1}, \quad\quad G_0^\dagger = G_0.
\label{eq:ghermite}
\end{equation}
Upon using the conjugations, the norm of the state $\vert \zeta ; k\rangle
:=((-)^{\nu N + \mu}G_{-1})^k \vert \zeta; 0 \rangle / k!$ is
\begin{equation}
\parallel \vert\zeta ; k \rangle \parallel^2 =
  ((-)^{\nu N+\mu})^k \left( \begin{array}{c}  2\zeta +k-1 \\ k
                        \end{array} \right). \label{eq:gnorm}
\end{equation}
The Hermitian conjugations  and the signature of the norm depend on the
value of $\nu N+\mu$.
When $\nu N+\mu$ is even,  eqs.(\ref{eq:ghermite},\ref{eq:gnorm})
say $U_k$ is the classical $su(1,1)$
and $V_k$ is the unitary infinite dimensional $su(1,1)$-module.
On the other hand, when $\nu N+\mu$ is odd, $U_k$ is the classical $su(2)$ and
$V_k$ is a non-unitary infinite dimensional $su(2)$-module.
Hereafter we write $V_k$ as $V_\zeta^{cl}$.

We turn to $U_r$ and $V_r$.
Let us define $\mho_j$ to be the finite dimensional module by rewriting the
level $r=0,1,\cdots,2j$ in $V_r$ by means of $m=-j+r$, that is
\begin{equation}
\mho_j= \{\vert j; m \rangle \;\vert\; \vert j; m\rangle :=
\frac{((-)^{\nu+1}J_-)^{j-m}}{[j-m]!}\vert j; j\rangle, \quad m=-j,
\cdots, j \}  \label{eq:defjmodule}
\end{equation}
Instead of (\ref{eq:defxp})-(\ref{eq:defk}), the actions of $J_\pm,\, \K$
on the module $\mho_j$ are written as
\begin{equation}
J_\pm \vert j; m\rangle =(-)^{\nu+1}[j\pm m+1] \vert j; m\pm 1\rangle,
 \quad \K\vert j; m\rangle =q^m \vert j; m\rangle
\label{eq:jaction}
\end{equation}
with $J_+\vert j; j\rangle=J_- \vert j; -j\rangle =0$.
Upon using (\ref{eq:jaction}), one finds the relations among
$J_\pm, \K$ to be
\begin{equation}
[J_+, J_-] = \frac{\K^2 - \K^{-2}}{q-q^{-1}}, \quad
\K J_\pm = q^{\pm1}J_\pm \K,
\label{eq:jrelation}
\end{equation}
and Hermitian conjugations are defined by

\begin{equation}
     J_\pm^\dagger = (-)^{\nu+1} J_\mp, \quad\quad \K^\dagger =\K^{-1}.
\label{eq:jhermite}
\end{equation}
With the Hermitian conjugations and the definition of $\vert j; m\rangle$
given in (\ref{eq:defjmodule}), the norm of the state
$\vert j ; m \rangle$ can be calculated as
\begin{equation}
\parallel\vert j; m \rangle \parallel^2 = ((-)^{\nu+1})^{j-m}
            \left[ \begin{array}{c} 2j \\ j-m \end{array} \right]_q
\label{eq:jnorm}
\end{equation}
Notice that the $q$-binomial coefficient in eq.(\ref{eq:jnorm}) is always
positive because $[x]=\frac{\sin (\pi x/N)}{\sin (\pi /N)}>0$ for $x<N$.
Relation (\ref{eq:jrelation}) and the conjugation (\ref{eq:jhermite})
together with the norm (\ref{eq:jnorm}) say that when
$\nu+1$ is even, $U_r$ is $U_q(su(2))$ and $\mho_j$ is a $(2j+1)$-dimensional
unitary $U_q(su(2))$-module, whereas  when $\nu +1$ is odd $U_r$ is
$U_q(su(1,1))$ and $\mho_j$ is a $(2j+1)$-dimensional non-unitary
$U_q(su(1,1))$-module.

Now we have understood the two spaces $V_\zeta^{cl}$ and $\mho_j$ together
with the unitarity conditions of them. We then wish to ask how the unitarity
of the original module $\V$ relates to the unitarity of $V_\zeta^{cl}$ and
$\mho_j$.  It is easy to answer the question by noticing that
the norm of the state $\vert h_{\mu\nu} ; kN+r \rangle$ is written as

\begin{equation}
\parallel \vert h_{\mu\nu} ; kN+r \rangle \parallel^2 =\,
 \parallel \vert j; m \rangle \parallel^2 \cdot
             \parallel \vert \zeta ; k \rangle \parallel^2.
\end{equation}

\noindent
Therefore $\V$ is unitary if and only if both $V_\zeta^{cl}$ and $\mho_j$ are
unitary and we have shown the following proposition:

\vspace{.5cm}
\noindent
{\bf Proposition 3.2}\quad
The irreducible highest weight module $\V$ is unitary if and only if
\begin{equation}
      \nu \in 2{\Bbb N}-1, \quad \mbox{and}\quad
     \mu\in \left\{  \begin{array}{cl}  2{\Bbb N}-1 &
                          \mbox{if}\quad N\in 2{\Bbb N}-1  \\
                        2{\Bbb N} & \mbox{if}\quad  N\in 2{\Bbb N}
                             \end{array} \right.
\end{equation}

\vspace{.5cm}
\noindent
Now the proof of Theorem 3.1 has been completed.

We end this section with presenting irreducible decomposition of the
completely reducible $\ncom$-module ${\bf V}^{su_q(1,1)}$.
Basically ${\bf V}^{su_q(1,1)}$ is a direct sum of $\V\simeq V_\zeta^{cl}
\otimes \mho_j$ and the values $\zeta$ and $j$ take
$\zeta\in{\Bbb N}/2$ and
$j\in\{0, \frac{1}{2}, 1, \frac{3}{2}, \cdots, \frac{N-1}{2}\}$.
However the maximal value of $j,\;i.e.,\: j=(N-1)/2$ is problematic
as in the $\com$ case.
In $\com$ with $q=\exp\pi i/N$, the highest weight state is restricted
such as $\vert j; j \rangle \in {\rm Ker}\,J_+ / {\rm Im}\,J_+^{N-1}$
\cite{PS}. Hence the value $j=(N-1)/2$ is excluded.
The quantum dimension defined by ${\rm Tr}_{R_j} \K^2$ for the highest weight
representation $R_j$ of $\com$ is zero for $R_{N-1/2}$.
In our case, using the character formula (\ref{eq:character}),
the quantum dimension is calculated as follows;
\begin{equation}
d_q:={\rm Tr}_{\V}K^2 = \chi_\zeta\cdot\chi_j,
\end{equation}
where
\begin{equation}
\chi_\zeta= \frac{q^{2N\zeta}}{1-q^{2N}}, \quad
\chi_j = \frac{q^{2j+1}-q^{-2j-1}}{q-q^{-1}}.
\end{equation}
When $j=(N-1)/2$, $\chi_j=0$ as in the $\com$ case, but $d_q$ is finite.
In contrast, when $j\le (N-2)/2$, $\chi_j = [2j+1]$ and the quantum dimension
$d_q$ diverges as expected. Therefore we have shown that
\begin{equation}
{\bf V}^{su_q(1,1)} = \left(\bigoplus_{\zeta\in{\bf N}/2}V_\zeta^{cl}\right)
\otimes \left(\bigoplus_{j\in{\cal A}}\mho_j\right),
\end{equation}
where ${\cal A}=\{0, \frac{1}{2}, 1, \frac{3}{2}, \cdots, \frac{N-2}{2}\}$.


\subsection{Clebsch-Gordan Decomposition}

Let us study the Clebsch-Gordan (CG) decomposition for the tensor product of
two irreducible highest weight representations of $\ncom$,
\begin{equation}
V_1 \otimes V_2\;\longrightarrow\; V_3,
\end{equation}
where $V_i:=V_{\mu_i, \nu_i}^{irr}\simeq V_{\zeta_i}^{cl}\otimes \mho_{j_i}$.
The quantum CG coefficient, known as the $q$-$3j$ symbol, for $\ncom$ has been
obtained in Refs.\cite{LK,Ai} when $q$ is generic.
In this section, in order to derive the CG decomposition rule for $\ncom$
with $\q$ we will make full use of the result obtained by Liskova and Kirillov
\cite{LK}.   The CG coefficient they have given is,
\begin{eqnarray}
&&\left[ \begin{array}{ccc} h_1 & h_2 & h_3 \\
                             M_1 & M_2 & M_3 \end{array}\right]^{su_q(1,1)}_q
=C(q)\delta_{M_1+M_2,\,M_3}\tilde{\Delta}(h_1, h_2, h_3)  \nonumber \\
&&\times \left\{
\frac{[2j-1][M_3-h_3]![M_1-h_1]![M_2-h_2]![M_1+h_1-1]![M_2+h_2-1]!}
      {[M_3+h_3-1]!}\right\}^{1/2} \nonumber \\
&&\times\sum_{R\ge 0}(-)^R q^{\frac{R}{2}(M_3+h_3-1)}
                \frac{1}{[R]![M_3-h_3-R]![M_1-h_1-R]![M_1+h_1-R-1]!}
\nonumber\\
&&\quad\cdot\frac{1}{[h_3-h_1-M_1+R]![h_3+h_2-M_1+R-1]!},
\end{eqnarray}
where $C(q)$ is a factor which is not important for our analysis below and
\begin{eqnarray}
&{}&\tilde{\Delta}(h_1, h_2, h_3)  \\
&{}&\quad =\{[h_3-h_1-h_2]![h_3-h_1+h_2-1]!
                        [h_3+h_1-h_2-1]![h_1+h_2+h_3-2]!\}^{1/2}. \nonumber
\end{eqnarray}
The notations $h_i=h_{\mu_i\,\nu_i}=\zeta_i N-j_i, \;
M_i=h_i+k_iN+r_i=(\zeta_i+k_i)N-m_i$ have been made.
Of course for our case, $i.e.$, $\q$, the CG coefficient is not necessarily
well-defined due to the factor $[N]=0$.
What we will do is to count the number of $[N]$ and look for the condition
such that the numbers of $[N]$ in the numerator and in the denominator should
be equal.
Lengthy examination derives the following result: Finite CG coefficients
exist if and only if
\begin{equation}
\begin{array}{l}
\zeta_1 + \zeta_2 \le \zeta_3, \\
{}\\
\vert j_1 - j_2 \vert -1 < j_3 \le {\rm min}\,(j_1+j_2, N-2-j_1-j_2).
\end{array}
\end{equation}
It should be noticed that, on the module $\V$, the coproduct of the operators
$K$ and $L_0$ are $\Delta(K)=K\otimes K$ and $\Delta(L_0)=L_0\otimes 1 +
1\otimes L_0$. The coproduct of the Cartan operator yields the conservation
law of the highest weights, physically speaking, the conservation of the
spin or angular-momentum along the $z$-axis.
Now, from the above coproducts, we have the conservation laws
$(\zeta_1+k_1) + (\zeta_2+k_2) = (\zeta_3+k_3)$ and $m_1 + m_2 = m_3$
(mod $N$).
Therefore the minimum value of $j_3$ is just $\vert j_1-j_2 \vert$ because
the difference between $\vert j_1-j_2 \vert$ and $j_3$ is always integer.
We then obtain the following decomposition rule of the tensor product of
two representations of $\ncom$,
\begin{equation}
\left( V_{\zeta_1}^{cl}\otimes \mho_{j_1}\right) \otimes
\left( V_{\zeta_2}^{cl}\otimes \mho_{j_2}\right)
= \left( \bigoplus_{\zeta_1+\zeta_2\le\zeta_3}V_{\zeta_3}^{cl}\right)
\otimes \left(\bigoplus_{j_3=\vert j_1-j_2\vert}^{{\rm min}\,
\{j_1+j_2, N-2-J_1-j_2\}}\mho_{j_3}\right).
\end{equation}
The decomposition rules for the tensor products of $V_\zeta^{cl}$ and of
$\mho_j$ are the same as those for the tensor products of the non-compact
representations of $\slc$ and of the compact
representation of $U_q(\slc)$, respectively.
However, taking the unitarity conditions for $V_\zeta^{cl}$ and $\mho_j$ and
the conservation laws of the highest weights into account,
we find that all the classical modules $V_{\zeta_i}^{cl},\,(i=1,2,3)$ cannot
be simultaneously unitary, and similarly for all the modules
$\mho_{j_i}, \, (i=1,2,3)$.


\subsection{Representation of $SL(2,{\bf R}) \otimes U_q(su(2))$ }

Through this section we suppose that the $\ncom$-module $\V$ is unitary.
Namely, the integers $\mu$ and $\nu$ take their values in accordance with
proposition 3.2.
Now we have found the classical sector $\slr\simeq su(1,1)$ in $U_q(su(1,1))$
and, therefore,  we can construct a representation of
$\SL\otimes U_q(su(2))$ via the representations we have obtained.

The basic strategy we will follow is to represent the $\SL$ sector,
roughly speaking,  as follows:
Let $G$ be a semi-simple Lie group, and $T$ a maximal torus.
The homogeneous space  $G/T$ has a complex homogeneous structure,
$i.e.$, the group $G$ acts on $G/T$ by means of holomorphic
transformation.
Then we can interpret the unitary irreducible representations of
$G$ as spaces of holomorphic sections of holomorphic line bundles
over $G/T$.  In our case, the group is $G=\SL$ and $T=U(1)$.
The homogeneous space $D=\SL/U(1)$ can be identified with
the complex upper half plane or alternatively with the Poincar\'e disk
$\vert w \vert < 1$.
Now we wish to obtain a representation not only of $\SL$
but also of $U_q(su(2))$, that is to say, the holomorphic sections are also
the representations of the quantum Lie algebra $U_q(su(2))$.
This means that each section should have the index with respect to
$U_q(su(2))$ as well.

Let $L$ be a line bundle over $\SL/U(1)$, and define
$L_j:=L\otimes \mho_j$.
We then construct $\Psi_{j,m}^\zeta\in L_j$, that is, $\Psi_{j,m}^\zeta$
is a holomorphic section of $L$ and also a vector in $\mho_j$.
Let ${\cal H}$ be the Hilbert space spanned by $\zk$.
Having an element $\langle \Psi \vert \in {\cal H}^\dagger$,
we give  the section by,
\begin{equation}
\Psi_{j,m}^\zeta(w):=\langle \Psi \vert \sum_{k=0}^\infty w^k
\vert \zeta; k \rangle
\otimes \vert j; m \rangle,
\end{equation}
Under the action of $\SL$, $w\,\rightarrow\, w'=  (aw+b)/(cw+d)$,
$\fun$ transforms as
\begin{equation}
\fun\;\longrightarrow\; {\Psi'}^\zeta_{j,m}(w')=
\left(\frac{1}{cw+d}\right)^{2\zeta}
\Psi^\zeta_{j,m}\left( \frac{aw+b}{cw+d} \right)  \label{eq:sectrans}
\end{equation}
Note that the action of $t\in U(1)$ is
$t\cdot\fun=\pi(t)\fun$ with $\pi(t)\in {\Bbb C}$.
It is well known that, on $L$,  we can pick a hermitian  metric
$(f, g)_\zeta= e^\eta f^*\cdot g$ by choosing the symplectic (K\"{a}hlar)
potential $\eta=2\zeta\log(1-\vert w\vert^2)$, $i.e.$, the curvature form on
$L$ is given by $w=-i \bar{\partial}\partial \eta(w^*,w)$.
With this hermitian metric,
the inner product on the space of sections of $L$ is given by
\begin{equation}
\langle f, g \rangle_\zeta = \int_{\vert w\vert < 1}
d\mu\,e^{\eta(w^*,w)} f^*\cdot g.
\end{equation}
where $d\mu =(2\zeta - 1)/\pi dw^* dw(1-\vert w\vert^2)^{-2}$ is the $\SL$
invariant measure.
Hence the inner product on $L_i$ is
\begin{eqnarray}
&{}&\langle \Psi^\zeta_{j,m}, \Psi^\zeta_{j,m'} \rangle_\zeta \nonumber \\
&{}&=\delta_{m,m'}\,
        \left[\begin{array}{c} 2j \\ j-m \end{array}\right]_q
  \frac{2\zeta -1}{\pi} \int_{\vert w\vert < 1} dw^* dw
(1-\vert w \vert^2)^{2\zeta-2}\psi_\zeta^*(w)\cdot \psi_\zeta(w),
\end{eqnarray}
where $\psi^\zeta(w)=\langle\Psi\vert \sum_{k=0}^\infty w^k \zk$.
It is worthwhile to notice that from equaton (\ref{eq:sectrans}),
the group $\SL$ now acts in a single valued way because $2\zeta\in{\Bbb N}$.
In the classical theory of representation, the condition that the $\SL$-spin
$\zeta$ should be a half-integer stems from the requirment of the group action
to be single-valued, while in the representation theory of
$U_q(su(1,1))$ at a root of unity,  the condition originates from the
requirment that all states have definite norms.
The action of the Lie algebra $\slr$ on $\fun$ is obtained upon
defining the action of $g\in\slr$ on the sections of $L$
as $g\cdot\psi^\zeta(w)=\langle\Psi\vert \sum_{k=0}^\infty w^k g\zk$.
By using $G_n$ actions (\ref{eq:deflp})-(\ref{eq:deflz}), we obtain

\begin{equation}
\hat G_n \fun= \left( w^{n+1} \partial_w + \zeta(n+1)w^n \right)\fun,
\quad\quad n= 0, \pm1, \label{eq:hatgaction}
\end{equation}

\noindent
where we have denoted the $\slr$-actions on the space of holomorphic
vectors as $\hat G_n$.
The right hand side of (\ref{eq:hatgaction}) spans infinitesimal
transformations of $\SL$, $w\rightarrow w+\epsilon(w)$
with $\epsilon(w)=\alpha w^2+\beta w +\gamma$.

On the other hand, the holomorphic vector $\Psi^\zeta_{j,m}$ transforms
under $\com$ as follows;
\begin{equation}
\begin{array}{l}
     J_\pm \Psi_{j,m}^\zeta(w) =
                    [j \pm m + 1] \Psi_{j,m\pm 1}^\zeta(w), \\
{}\\
    {\cal K} \Psi_{j,m}^\zeta(w) = q^m \Psi_{j,m}^\zeta(w),
\end{array}
\end{equation}
with  $ J_+\Psi_{j,j}^\zeta(w) = J_-\Psi_{j,-j}^\zeta(w) = 0$.
In the above we have given only the highest weight representations for
the $\com$ sector. However, it is possible to represent this sector
in other ways which are more useful for physical applications.
In particular, the construction given in \cite{GS,RRR} of the representation
space of $\com$ is important for the further investigations,
especially the connections with 2$D$ conformal field theories.

Let us turn our attention to  another remarkable feature of $\ncom$ at a
root of unity, that is, the connection with the generalized supersymmetry.
Of course we can easily guess this connection from (\ref{eq:ldef}) together
with the fact that $L_{0,\pm1}$ generate $\slr$ when the representation is
unitary.
In the following we will explicitly show that the operators $L_{0,\pm1}$ in
(\ref{eq:ldef}) can be written as the infinitesimal transformations of the
Poincar\'e disk and therefore the operators $X_\pm$ can be interpreted as the
$N$-th roots of the transformations.
Our discussion proceeds as follows.
First of all, let us find suitable functions on the disk so that
the operators $L_{0,\pm1}$ act on them as the infinitesimal transformations.
In the previous part of this section, we constructed the holomorphic vector
$\fun$ by summing $\zk\otimes\jm$ over the level $k$ in the $\slr$ sector.
Noticing that on $\fun$, $\hat G_n$ rather than $L_n$ played the roles of
such transformations, we should change our standing point and obtain other
holomorphic vectors on which $L_n$ naturally act.
Therefore, in this case, we have to construct the vectors by means of the
original states $\hkr\in V_{\mu,\nu}$ instead of $\zk\otimes\jm$.
Notice here that we do not restrict the highest weight modules to the
irreducible modules and so the level $r$ runs from $0$ to $N-1$.
Let us define the holomorphic vectors as follows;

\begin{equation}
\pfun_r(w) :=\langle\Phi\vert \sum_{k \in {\Bbb Z}_+}(-)^{\frac{1}{2}k(k-1)+kr}
          w^k \hkr,   \label{eq:defphi}
\end{equation}

\noindent
where $\vert w\vert < 1$. $\pfun_r(w)$ also behaves as a holomorphic section
of the line bundle over $\SL/U(1)$.
Now we have $N$ holomorphic functions
$\pfun_0(w),\,\pfun_1(w),\cdots, \pfun_{N-1}(w)$.
On the contrary, we had $2j+1$ functions $\Psi^\zeta_{jj}, \Psi^\zeta_{jj-1},
\cdots, \Psi^\zeta_{j-j}$ in the previous case.
{}From eq.(\ref{eq:qxpaction}) one can easily calculate the actions of
$X_+$, denoted as $\hat X_+$, on these functions as follows;
\begin{equation}
\hxp^\vp\pfun_r(w)=\left\{\begin{array}{ll}
       {[r+1]} \pfun_{r+1}(w), & \quad{\rm for}\quad 0\leq r \leq N-2 \\
   \rule{0mm}{.7cm}
       {[N]}_\vp\partial_w \pfun_0(w), & \quad{\rm for}\quad r=N-1.
                   \end{array}\right.
\label{eq:hxpact}
\end{equation}

\noindent
Similarly upon using eq.(\ref{eq:qxmaction}) we obtain
\begin{equation}
-\hxm^\vp\pfun_r(w)=\left\{\begin{array}{ll}
  w{[\mu]} \pfun_{N-1}(w), & \quad{\rm for}\quad r=0 \\
     \rule{0mm}{.5cm}
  {[\mu-r]} \pfun_{r-1}(w), & \quad{\rm for}\quad 1\leq r\leq \mu-1 \\
     \rule{0mm}{.5cm}
  {[N]}_\vp(w\partial_w +2\zeta)\pfun_{\mu-1}(w),
                                          &\quad{\rm for}\quad r=\mu \\
     \rule{0mm}{.5cm}
  {[N+\mu-r]} \pfun_{r-1}(w), & \quad{\rm for}\quad \mu+1\leq r\leq N-1.
                   \end{array}\right.
\label{eq:hxmact}
\end{equation}

\noindent
Here we have introduced the symbols $[N]_\vp$ and $\hxpm^\vp$
such that $[N]_\vp \neq 0$ and $\lim_{\vp\rightarrow 0} [N]_\vp = [N]=0$,
and $\lim_{\vp\rightarrow 0} \hxpm^\vp = \hxpm$.
Now let us calculate the $N$-th powers of $\hxpm^\vp$. We obtain

\begin{eqnarray}
 \lim_{\vp\rightarrow0}\,  \frac{(\hxp^\vp)^N}{[N]_\vp!} \pfun_r(w) &=&
   \partial_w\pfun_r(w)  \label{eq:XNpact} \\
 \lim_{\vp\rightarrow0}\,  \frac{(-\hxm^\vp)^N}{[N]_\vp!} \pfun_r(w) &=&
  \left\{\begin{array}{lc}
      \left(w^2\partial_w +2\zeta w\right)\pfun_r(w), & 0\leq r\leq \mu-1 \\
     \rule{0mm}{.5cm}
      \left(w^2\partial_w +2(\zeta+\frac{1}{2}) w\right)\pfun_r(w), &
                                                        \mu\leq r\leq N-1
        \end{array}\right.
     \label{eq:XNmact}
\end{eqnarray}

\newcommand{\wt}{\widetilde}

\noindent
Thus we  have shown that
$\hxpm^\vp$, which move the function $\pfun_r(w)$ to $\pfun_{r\pm 1}(w)$
accoeding to (\ref{eq:hxpact},\ref{eq:hxmact}), are related to the
$N$-th roots of infinitesimal transformations of the Poincar\'e disk.
Furthermore, eq.(\ref{eq:XNmact}) tells us that the functions $\pfun_r(w)$
for $0\leq r\leq \mu-1$ have dimensions $\zeta$ and the functions $\pfun_r(w)$
for $\mu \leq r\leq N-1$ have dimensions $\zeta+\frac{1}{2}$.
Let $\wt{\Phi}_r^\zeta(w)$ be the former functions, {\em i.e.}, $\pfun_r$
with dimensions $\zeta$  and $\Xi_r^{\zeta+\frac{1}{2}}(w)$ be the
latter with dimensions $\zeta+\frac{1}{2}$.
In the above, we examined $\hxpm$ and the $N$-powers of them only.
Of course we can also obtain
\begin{equation}
\lim_{\vp\rightarrow0}\, \frac{1}{2}
      \left[ \begin{array}{c} 2\hat{H}+N-1 \\ N \end{array} \right]_q
      = (w\partial_w + h ). \label{eq:znm}
\end{equation}

\noindent
where $\hat K=q^{\hat H}$ and $h$ is the dimension, {\em i.e.}
$h=\zeta$ for the functions $\wt{\Phi}^\zeta_r(w)$ and $h=\zeta+\frac{1}{2}$
for the functions $\Xi^{\zeta+\frac{1}{2}}_r(w)$.
Thus we have obtained all the generators which span the infinitesimal
holomorphic transformations of the homogeneous space $\SL/U(1)$
and obtained two kinds of functions.  One is the set of
functions whose dimensions are $\zeta$ and the other is the set of functions
whose dimensions are $\zeta+\frac{1}{2}$.
Moreover the generators $\hxpm$ mix between them.
Now we can conclude that $\ncom$ with the deformation parameter $\q$
may be viewed in terms of a $Z_N$-graded supersymmetry with the
upper half plane or Poincar\'e disk  interpreted as an external space.

We have constructed holomorphic vectors over the Poincar\'e disk
in two ways and obtained two sets, $\fun$ and
$\pfun_r(w)=\{\wt{\Phi}_r^\zeta(w), \Xi^{\zeta+\frac{1}{2}}_r(w)\}$.
We end this section with the discussion of the connection between them.
Upon the substitution $m=-j+r$, the actions of $\hxpm$ on
$\wt{\Phi}^\zeta_r(w)$ coincide with those of $J_\pm$ on $\fun$ except
the actions $\hxp \wt{\Phi}^\zeta_\mu$ and $\hxm \wt{\Phi}^\zeta_0$
corresponding to $J_+\Psi^\zeta_{jj}$ and $J_-\Psi^\zeta_{j-j}$,
respectively.
The latter vanish because  $\Psi^\zeta_{jj}$ is the
highest weight vector and $\Psi^\zeta_{j-j}$ is the lowest weight vector
with respect to $\com$.
However $\hxp \wt{\Phi}^\zeta_\mu$ and $\hxm \wt{\Phi}^\zeta_0$ do not
vanish but yield the functions $\Xi^{\zeta+\frac{1}{2}}_{\mu+1}$ and
$\Xi^{\zeta+\frac{1}{2}}_{N-1}$.
In other words, through these two actions the two classses of functions,
$\wt{\Phi}^\zeta_r$ and $\Xi^{\zeta+\frac{1}{2}}_r$, mix with each other after
taking the limit $\vp\rightarrow0$ in eqs.(\ref{eq:hxpact},\ref{eq:hxmact}).
Noticing that the functions $\Xi^{\zeta+\frac{1}{2}}_r$ have zero norms
because they correspond to the states lying between the $(kN+\mu)$-th level
to the $(kN+N-1)$-th level in the original module $V_{\mu,\nu}$,
we see that they are proportional to
$\sqrt{[N]_\vp}$ and may rescale them as
$\Xi^{\zeta+\frac{1}{2}}_r=\sqrt{[N]_\vp}\,\wt{\Xi}^{\zeta+\frac{1}{2}}_r$.
By the rescaling,  the set of functions $\wt{\Xi}^{\zeta+\frac{1}{2}}_r$
completely decouples from the set $\wt{\Phi}^\zeta_r$
after taking the limit $\vp\rightarrow0$.
We can then identify $\wt{\Phi}^\zeta_r(w)$ with $\fun$, and we have
another set of functions $\wt{\Xi}^{\zeta+\frac{1}{2}}_r$ whose dimensions
differ by $1/2$ from those of $\wt{\Phi}^\zeta_r$.

\vspace{0.7cm}


\section{$Osp(1\vert 2)$ and $U_q(su(1,1))$ with $q^2=-1$}

In this section we devote ourselves only to the case $N=2$, that is,
$[2]=0$ and find that $Osp(1\vert 2)$ can be represented in terms
of the representation of $U_q(su(1,1))$.
To this end it is convenient to introduce operators $\lp$ and $\lm$,
which are related to $X_\pm$ and $K$ by the relations
\begin{equation}
\lp = iq^{-\frac{1}{2}}KX_-, \quad\quad \lm = iq^{-\frac{1}{2}}X_+K.
\end{equation}
Further we define vectors $\phi_r^\zeta(w), \; r= 0, 1$ by means of the
highest weight representations of $U_q(su(1,1))$ by
\begin{equation}
\phi_r^\zeta(w) = \sum_{k=0}^\infty w^k \vert h_{\mu\nu};
            2k+r ), \quad\quad r=0, 1
\end{equation}
where we have introduced new weight vectors
\begin{equation}
  \vert h_{\mu\nu}; r ) := \frac{\lm^r}{\langle r \rangle !}
  \vert h_{\mu\nu}; 0 \rangle,
\end{equation}
with ${\langle r \rangle }=q^{r-1}{[r]}$.
The new weight vector $\vert h_{\mu\nu}; r )$ coincides with the original
one $\vert h_{\mu\nu}; r\rangle$ up to a phase factor.
We can, therefore, deal with the highest weight modules spanned by the new
weight vectors in the same fashion as  $V_{\mu,\nu}$.
In particular, the operator $\lp$ and $\lm$ act as
\begin{equation}
    \begin{array}{l}
      \lp \vert h_{\mu\nu}; r ) =  {\langle 2h_{\mu\nu}+r-1 \rangle }
      \vert h_{\mu\nu}; r-1 ), \\
  \rule{0mm}{.5cm}
      \lm \vert h_{\mu\nu}; r ) =  {\langle r+1 \rangle }
      \vert h_{\mu\nu}; r+1 ).
    \end{array}
\end{equation}
{}From these actions it is easily seen that the $\mu$-th state  has
zero norm as in the previous case.
Indeed, in $N=2$ case, the highest weight is given by (see eq.(\ref{eq:hmn})),
\begin{equation}
h_{\mu\nu}=\frac{1}{2}(2\nu - \mu +1). \label{eq:hmnt}
\end{equation}
Because $1\le\mu\le N$, there are two cases, $\mu=1$ and $\mu=2$;
in the case when $\mu=1$, $\po$ has zero norm, while neither $\pz$ nor
$\po$ has zero norm in the case when $\mu=2$.
We treat these cases separately.

We first examine the case when $\mu=1$.
The highest weight is given by $h_{1\nu}=\nu$.
Upon using the relation  $\langle 2n \rangle=\langle 2 \rangle n$
for an integer $n$,
the actions of $\lm$  on the vectors $\pz, \; \po$
are easily obtained,
\begin{equation}
  \lm\pz = \po, \quad\quad
  \lm\po = \langle 2 \rangle \partial_w \pz, \label{eq:ospp}
\end{equation}
and $\lp$ acts on them as
\begin{equation}
  \lp\pz = w \po, \quad\quad
  \lp\po = \langle 2 \rangle ( \nu+ w\partial_w) \pz. \label{eq:ospm}
\end{equation}
At first sight of eqs.(\ref{eq:ospp}) and (\ref{eq:ospm}) we might suspect
that $\pz$ and $\po$ are superpartner with each other and ${\cal L}_{\pm1}$
are the generators of supersymmetry transformations.
However we must be more cautious because the actions of
${\cal L}_{\pm1}$ on $\po$ are zero-actions due to the factor
$\langle2\rangle$.
Note that the vector $\po$ must be proportional
to $\sqrt{\langle 2 \rangle }$ since the norm of the vector is proportional to
$\langle 2 \rangle $.
Fortunately we can remedy the situation by scaling the operators and the
vector $\po$ as follows;
\begin{equation}
\begin{array}{l}
\phi^\zeta(w) = \pz,\quad\quad  \sqrt{\langle 2 \rangle }\,
\psi^\zeta(w) = \po, \\
\sqrt{\langle 2 \rangle }\,{\cal G}_{-1}
= \lm, \quad\quad \sqrt{\langle 2 \rangle }\,{\cal G}_1 = \lp.
\end{array}
\label{eq:scaling}
\end{equation}
The actions of ${\cal G}_\pm$ on $\phi^\zeta$ and $\psi^\zeta$ are
now non-vanishing and $\psi^\zeta$ has definite norm.

We are ready to discuss the connection between two-dimensional
supersymmetry and $U_q(su(1,1))$.
Let us define an infinitesimal transformation $\de$ as
\begin{equation}
\de:= a\,{\cal G}_1 + b\,{\cal G}_{-1},
\end{equation}
where $a, b$ are infinitesimal Grassmann numbers.
Under the transformation the fields $\ph$ and $\ps$  transform into
each other according to
\begin{equation}
   \begin{array}{l}
     \de\ph = \varepsilon(w)\ps, \\
  \rule{0mm}{.5cm}
     \de\ps = \left( \nu(\partial_w \vp(w)) + \vp(w) \partial_w \right) \ph,
   \end{array}
\end{equation}
where  $\vp(w)=aw +b$ is an anticommuting analytic function which
parametrises infinitesimal holomorphic transformation.
The commutation relations of two transformations
$\deo$ and $\dew$ are
\begin{equation}
   \begin{array}{l}
      {[\deo, \dew]}\ph = \left( \zeta (\partial_w \xi(w))
        +\xi(w)\partial_w \right) \ph,  \\
  \rule{0mm}{.5cm}
    {[\deo, \dew]}\ps = \left( (\zeta+\frac{1}{2})(\partial_w \xi(w))
        +\xi(w)\partial_w \right) \ps,
   \end{array}
\label{eq:sltr}
\end{equation}
with $\xi(w)=2\vp_1(w)\vp_2(w)$.
The right hand sides of equations (\ref{eq:sltr}) are just the
transformations of the fields $\ph$ and $\ps$ having dimensions $\zeta$ and
$\zeta+\frac{1}{2}$, respectively, under the the infinitesimal transformation
of $\SL$,  $w\,\rightarrow\,w+\xi(w)$.
We can therefore conclude that the infinitesimal transformation
$\de$ which is written in terms of generators of $U_q(su(1,1))$ is
just the \lq square root' of infinitesimal $SL(2,{\bf R})$
transformation, that is to say, $\de$ is an infinitesimal supersymmetry
transformation.
Further the fields $\ph$ and $\ps$ which are constructed in terms of
the highest weight representations of $U_q(su(1,1))$ can be regarded as
superpartners with each other.
Finally, $Osp(1\vert 2)$ algebra is obtained as follows:
Let $L_{\pm1}=({\cal G}_{\pm1})^2$ and $F_{\pm\frac{1}{2}}={\cal G}_{\pm1}$,
then the following commutation relation are easily checked on the fields
$\phi^\zeta(w)$ and $\psi^\zeta(w)$ to be
\begin{equation}
     \begin{array}{ll}
    {[L_n, L_m]}=(n-m)L_{n+m},  & n, m=0, \pm1, \\
    {[L_n, F_r]}=\left( \frac{1}{2}n-r \right)F_{n+r},
                                & r=\frac{1}{2}, -\frac{1}{2}, \\
    {\{F_r, F_s\}}= 2L_{r+s},     &  r, s =\frac{1}{2}, -\frac{1}{2}.
     \end{array}
\end{equation}
Thus we have succeeded in building the super-algebra $Osp(1\vert 2)$
and its representation in terms of the representation of
$U_q(su(1,1))$ when $q^2=-1$ and $\mu=1$, $i.e.$, all states at the
$(2{\Bbb N}-1)$-th levels are zero-norm states.

Next we turn to the $\mu=2$ case. By eq.(\ref{eq:hmnt}),
the highest weight is given by $\zeta-\frac{1}{2}$.
On the contrary to the case when $\mu=1$, no zero-norm state appears.
The actions of  $\lp$ and $\lm$ on $\phi_{0,1}$ are as follows;
\begin{equation}
\begin{array}{l}
  \lp\pz = \langle 2 \rangle (\nu w + w^2\partial_w)\po, \quad
  \lp\po = \pz,  \\  {}  \\
  \lm\pz = \po, \quad
  \lm\po = \langle 2 \rangle \partial_w \pz.
\end{array}
\end{equation}
Unfortunately, we cannot find any consistent ways to remove the factor
$\langle 2 \rangle$ as in the previous case (\ref{eq:scaling}).


\section{Discussion}

In this article the highest weight representations of $\ncom$ when
$\q$ has been  investigated in detail.
We have shown that the highest weight module $\V$ is isomorphic to the tensor
product of two highest weight modules $V_\zeta^{cl}$ and $\mho_j$.
This fact played a key role of this work, and novel features of $\ncom$
originated from this structure of $\V$. The module $\vc$ is a
classical non-compact $\slc$-module, while $\vq$ is a
$(2j+1)$-dimensional module of the quantum universal envelopping algebra
$U_q(\slc)$.  Theorem 3.1 states what $\vc$ and $\vq$  are.
In particular, when the original $\ncom$-module $\V$ is unitary,
$V_\zeta^{cl}$ is the unitary highest weight module of
$su(1,1)\simeq \slr$ and $\vq$ is the unitary highest weight $\com$-module.
In the following we restrict our discussions to this case, {\it i.e.},
$\V$ is unitary.

We summarize here the novel features of $\ncom$ when $\q$:
First we should notice that the non-compact nature appears only through the
classical module $\vc$, and the effects of $q$-deformation arise only from the
compact sector $\vq$.
Since the non-compact sector $\vc$ is classical, a representation of the Lie
group $\SL$ is naturally induced.
Indeed, we gave a representation of $\SL\otimes\com$ by means of the
holomorphic vector $\fun=\psi_\zeta(w)\otimes\vert j; m\rangle$.
Here we used holomorphic sections $\psi_\zeta(w)$ of a line bundle over the
homogeneous space $\SL/U(1)$ in order to represent the $\SL$ sector and
$\vert j; m\rangle\in\vq$ is a weight vector with respect to $\com$.
With our deformation parameter $q$, $i.e.$, $\q$, we have shown that
the value of the highest weight $j$ lies in ${\cal A}=\{0, \frac{1}{2}, 1,
\cdots, \frac{N-2}{2}\}$.
Notice that this finiteness of the number of
the highest weight states for the $\com$ sector comes from the
condition that the original highest weight representations $\V$ of
$\ncom$ be well-defined, that is, every state in them has finite norm.
The representation $\fun$ says that every point on the homogeneous space,
({\it i.e.}, the upper half plane or the Poincar\'e disk)
has the representation space of $\com$.
In this sense, we suggest that the non-compact homogeneous
space can be viewed as a base space or an external space and the
representation space of $\com$ as an internal space.

We have also discussed the connection between $\ncom$ with $\q$ and
$Z_N$-graded supersymmetry by presenting $N$ holomorphic vectors,
denoted as $\pfun_r(w),\,r=0, 1, \cdots, N-1$, in another way.
The generators, $X_\pm, K$ of $\ncom$ act on them and map $\pfun_r$ to
$\pfun_{r\pm1}$.
On the other hand, the operator $L_n$ which are related to the $N$-th
powers of $X_\pm, K$ by the relations (\ref{eq:ldef}) generate the holomorphic
transformations of the functions under the infinitesimal transformations
of the homogeneous space $\SL/U(1)$.
In this sense, we may say that generators of $\ncom$ give rise to
$Z_N$-graded supersymmetry transformations and the $N$-th powers of them
are related to the infinitesimal transformations with respect
to the external space.
Furthermore, by observing the transformations under $L_n$, we have shown
that these $N$ functions separate into two classes.
One of them is the set of functions,
$\wt{\Phi}^\zeta_r(w)$, $r=0\sim 2j$, with dimensions $\zeta$
and the other is the set of functions, $\Xi_r^{\zeta+\frac{1}{2}}(w)$,
$r=2j+1\sim N-1$, which have dimensions $\zeta+\frac{1}{2}$ and have
zero norms.
That is to say, the functions $\wt{\Phi}^\zeta_r$ and
$\Xi_r^{\zeta+\frac{1}{2}}$ behave as the covariant vectors with dimensions
$\zeta$ and $\zeta+\frac{1}{2}$, respectively, under $\slr$.
In particular, we have shown the explicit realization of two-dimensional
supersymmetry $Osp(1\vert2)$ via the representation of $\ncom$ when the
deformation parameter satisfies $q^2=-1$.

We have also discussed the Clebsch-Gordan decomposition for the tensor
product of two irreducible highest weight modules and found that the
decomposition rules for the two sectors $V_\zeta^{cl}$ and $\mho_j$ coincide
with those for the classical non-compact representations of $\slr$ and the
representations of $\com$.

\vspace{0.5cm}

Finally, we would like to future issues to be investigated.
As mentioned in chapter 1, it is quite interesting to expect the relationship
between $\ncom$ and topological $2D$ gravity coupled with RCFT.
In order to make this expectation come true, we have to find a good
representation space of $\ncom$ for which such a physical theory is
associated \cite{MS2}.
The $Z_N$ graded supersymmetry implies that the internal symmetry, $\com$,
is not independent of the base manifold $\Sigma_g$ but yields the deformation
of the metric of $\Sigma_g$ through the $N$-th powers of the action.

Second, it is also interesting to investigate geometical aspect of our result.
As for the geometical viewpoint of quantum groups,
it is widely expected that quantum groups will shed light on the concept
of ``quantum'' space-time.
In particular, quantum groups in the sense of $A_q(G)$,
the $q$-deformation of the functional ring over the group $G$,
rather than $U_q({\Got{g}})$ play the central role in the noncommutative
geometry initiated by Manin\cite{Ma}, and Wess and Zumino \cite{WZ}.
Further Wess and Zumino have studied a $q$-deformed quantum mechanics
in terms of the noncommutative differential geometry based on $A_q(G)$.
The phenomena observed in this paper suggest that by the quantization of the
Poincar\'e disk, a certain ``$q$-deformed space'' appears as a
($q$-deformed) fiber at each point on the disk which remains classical.
This observation is reminiscent of the result obtained in Ref.\cite{Su}.
Actually, in order to construct $q$-deformed mechanics,
a $q$-deformed phase space was introduced in \cite{Su} by attaching
an internal space at each point on the phase space of the classical
mechanics and all effects of $q$-deformation stemed only from the internal
space.

\vspace{1.0cm}

\noindent
I am grateful to Dr. T. Matsuzaki for fruitful collaborations and
discussions. I would also like to thank Dr. H. W. Braden for
carefully reading this manuscript and valuable comments.

\newpage


\end{document}